\documentclass[aps,10pt,twocolumn,superscriptaddress,showpacs,reprintnumbers,amsmath,amssymb,floatfix]{revtex4}
\usepackage{longtable}
\usepackage{lscape}
\usepackage{graphics}
\usepackage{graphicx}

\begin{document}
\title{ The magnetic properties of $^{\rm 177}$Hf and $^{\rm 180}$Hf in the strong coupling deformed model.}
\author{S.~Muto}
\affiliation {Neutron Science Laboratory, KEK, Tsukuba, Ibaraki 305-0801, Japan}
\author{N.~J.~Stone}
\affiliation {Department of Physics and Astronomy, University of Tennessee, Knoxville, TN 37996, USA} 
\affiliation{Department of Physics, University of Oxford, Oxford, OX1 3PU, United Kingdom}
\author{C.~R.~Bingham}
\affiliation {Department of Physics and Astronomy, University of Tennessee, Knoxville, TN 37996, USA}
\affiliation {Oak Ridge National Laboratory, Oak Ridge, TN 37831, USA} 
\author{J.~R.~Stone}
\affiliation {Department of Physics and Astronomy, University of Tennessee, Knoxville, TN 37996, USA} 
\affiliation{Department of Physics, University of Oxford, Oxford, OX1 3PU, United Kingdom}
\author{P.~M.~Walker}
\affiliation{Department of Physics, University of Surrey, Guildford, Surrey GU2 7XH, United Kingdom}
\author{G.~Audi}
\affiliation{CSNSM, F-91405 Orsay, France}
\author{C.~Gaulard}
\affiliation{CSNSM, F-91405 Orsay, France}
\author{U. K\"{o}ster}
\affiliation{Institut Laue Langevin, F-38042 Grenoble, France}
\author{J.~Nikolov}
\affiliation{Department of Physics, University of Novi Sad, 21000 Novi Sad, Serbia}
\author{K.~Nishimura}
\affiliation{Faculty of Engineering, Toyama University, Toyama 930, Japan}
\author{T.~Ohtsubo}
\affiliation{Department of Physics, Niigata University, Niigata 950-2181, Japan}
\author{Z.~Podolyak}
\affiliation{Department of Physics, University of Surrey, Guildford, Surrey GU2 7XH, United Kingdom}
\author{L.~Risegari}
\affiliation{CSNSM, F-91405 Orsay, France}
\author{G.~S.~Simpson}
\affiliation{LPSC, F-38026 Grenoble, France}
\author{M.~Veskovic}
\affiliation{Department of Physics, University of Novi Sad, 21000 Novi Sad, Serbia}
\author{W.~B.~Walters}
\affiliation{Department of Chemistry and Biochemistry, University of Maryland, College Park MD 207 40, USA}
\date{\today}
 
\begin{abstract}
This paper reports NMR measurements of the magnetic dipole moments of two high-K isomers, the 37/2$^-$, 51.4 m, 2740 keV state in $^{\rm 177}$Hf and the 8$^-$, 5.5 h, 1142 keV state in  $^{\rm 180}$Hf by the method of on-line nuclear orientation. Also included are results on the angular distributions of gamma transitions in the decay of the  $^{\rm 177}$Hf isotope. These yield high precision E2/M1 multipole mixing ratios for transitions in bands built on the 23/2$^+$, 1.1 s, isomer at 1315 keV and on the 9/2$^+$, 0.663 ns, isomer at 321 keV.  
The new results are discussed in the light of the recently reported finding of  systematic dependence of the behavior of the g$_{\rm R}$ parameter upon the quasi-proton and quasi-neutron make up of high-K isomeric states in this region.
\end{abstract}
\maketitle
\section{\label{intro}Introduction}

This paper reports measurement of the magnetic dipole moments of the 51.4 m, 2740 keV, 37/2$^-$ isomer in $^{\rm 177}$Hf and of the 5.5 h, 1142 keV, 8$^-$ isomer in $^{\rm 180}$Hf, using the method of NMR on oriented nuclei, at the on-line NICOLE facility, ISOLDE, CERN. In addition, measurements of the angular distribution of gamma transitions from the $^{\rm 177m2}$Hf isomer are analysed to give precise values of the E2/M1 mixing ratios in transitions in bands built upon the 23/2$^+$, 1.1 s, 1315 keV isomer and the 9/2$^+$, 321 keV state of this nucleus.

Accurate measurements of the electromagnetic properties of ground states, isomers and rotational band states in deformed nuclei yield parameters of the deformed nuclear model used to describe these nuclei. In the region between Yb and W ground states and isomers have close-to-constant deformation and the parameters of the model can be explored thoroughly. In this work the main focus is on isomeric state, band-head, magnetic dipole moments to yield values of the single particle g-factor, g$_{\rm K}$, with good precision. In addition, analysis of the angular distribution of mixed M1/E2 transitions in the decay of the isomers studied yields values of the mixing ratio $\delta$ which is related to the g-factors g$_{\rm K}$, and g$_{\rm R}$, and intrinsic quadrupole moment Q$_{\rm 0}$, of the band-head (see Eq.~\ref{delt}). 

Interest in obtaining the magnetic parameters g$_{\rm K}$ and g$_{\rm R}$ in these states derives from their involvement with ideas of superconductivity in nuclei. In the strong coupling model of well-deformed nuclei the influence of the quasi-particle state upon the band properties arises through the dependence of the pairing strength, and hence the pairing gap $\Delta$, upon the available orbitals close to the Fermi surface. In the deformed potential each state is doubly degenerate and is available for pair scattering if empty, however the occupation of a state by a single quasi-particle renders it unavailable to pair scattering. This is the process of blocking \cite{dracoulis1998,wu2011}. Since the pairing gap is determined by the number of pairs and the numbers of states between which they can scatter, increasing the number of quasi-particles increases blocking and reduces the gap $\Delta$.

These considerations apply separately to protons and neutrons. Where the rotational energy levels of the bands are concerned, the relevant model parameter is the total moment of inertia $\mathcal{I}_{\rm tot}$ = $\mathcal{I}_{\rm p}$ +$\mathcal{I}_{\rm n}$. The stronger the pairing associated with the state the more $\mathcal{I}_{\rm tot}$ is reduced compared to the classical rigid body value. Blocking of either proton or neutron orbitals weakens the pairing and increases $\mathcal{I}_{\rm tot}$. When electromagnetic properties of the band are considered, transition matrix elements, intensity ratios, multipole mixing ratios and the collective g-factor g$_{\rm R}$ show different sensitivity to proton and neutron pairing, and the effects of blocking, through their differing effective charges, crudely 1 for protons and 0 for neutrons. In this limit g$_{\rm R}$ is given by the ratio $\mathcal{I}_{\rm p}$/$\mathcal{I}_{\rm tot}$, which yields the familiar simple result g$_{\rm R}$ $\sim$ Z/A for rigid rotation of the whole nucleus. It follows from these basic ideas that we can expect any increase in, for example, neutron blocking, by reducing the neutron pairing gap and increasing the neutron contribution to $\mathcal{I}_{\rm tot}$, to reduce g$_{\rm R}$ relative to Z/A, whilst increased proton blocking will have the opposite result.

The new experimental results presented here, in combination with a much larger body of data from existing literature, have been used recently to explore these concepts \cite{stone2013}. It has been shown that, within the rather limited data presently available, the principle of additivity can be relied upon to give good estimates of the single particle g-factor, g$_{\rm K}$, in multi-quasi-particle isomers. The use of this principle to estimate g$_{\rm K}$ in many bands revealed a wide, systematic, dependence of the collective g-factor, g$_{\rm R}$, upon the neutron and proton quasi-particle make-up of the band-head states, presenting a new challenge to the theory of superconductivity, pairing and blocking in these deformed nuclei. This finding gives additional reason to extend, in particular, band-head magnetic moment measurements of good accuracy and to seek accurate mixing ratios to yield g$_{\rm R}$, as are reported here.

The theoretical context of this work is given in Sec.~\ref{theory}. Experimental details and new data on the angular properties of gamma transitions in the decay of the 37/2$^-$ K-isomer of $^{\rm 177}$Hf are described in Secs.~\ref{exp} and \ref{res} together with NMR results which yield the magnetic dipole moments of the 37/2$^-$ K-isomer and of the 8$^-$ isomer in $^{\rm 180}$Hf. Consistency of the new results with the findings of \cite{stone2013} is discussed in Sec.~\ref{dis}.

\section{\label{theory}Theoretical framework}
Relevant expressions from the strong coupling model concerning electromagnetic properties of the nuclear states and transitions are given here \cite{bm}. For the static magnetic dipole moment of a band state of spin I based on a band head of spin K 
\begin{equation}
\mu = g_{\rm R}I + (g_{\rm K} - g_{\rm R})\frac{K^{\rm 2}}{I + 1}
\label{mu}
\end{equation}
which, for the band head K = I, gives
\begin{equation}
\mu = g_{\rm K} [I^{\rm 2}/(I + 1)]  + g_{\rm R}[I/(I + 1)]
\label{gk} 						
\end{equation}
Since the values of spin I in this paper are large, this expression makes it clear that the moment has only weak dependence upon the `collectiveÕ g$_{\rm R}$ factor and is largely determined by the quasi-particle g$_{\rm K}$ factor.

 For a pure multi-quasi-particle state for which $K$ is the simple sum of the constituent quasi-particle K$_{\rm i}$'s, the quasi-particle g-factor g$_{\rm K}$ is given in terms of the constituent quasi-particle g$_{\rm Ki}$-factors by 
\begin{equation}
g_{\rm K} = \frac{1}{K}\sum_{i}K_{\rm i}g_{\rm K_{\rm i}}.						
\label{gks}		
\end{equation}
Two spectroscopic variables, the E2/M1 ratio $\delta$ in transitions from a level of spin I to a lower level of spin I - 1 and the branching ratio of transitions from a state of spin I to lower states of spin I - 1 and I - 2, depend upon the difference (g$_{\rm K}$ -  g$_{\rm R}$) of the two (collective and quasi-particle) g-factors. These variables can both be written in terms of $\delta$ given by \begin{equation}
\delta = \frac{0.933E_{\rm \gamma}Q_{\rm 0}}{(g_{\rm K} - g_{\rm R}) \sqrt{(I^{\rm 2} - 1)}},		
\label{delt}		
\end{equation}
where E$_{\rm \gamma}$ is the transition energy in MeV and Q$_{\rm 0}$ is the intrinsic quadrupole moment in $e$b. From these expressions it is seen that to separate the quasi-particle g-factor g$_{\rm K}$ from the rotation g-factor g$_{\rm R}$ it is necessary to have data on either transition branching ratios from states in the band to lower states of spin reduced by one and two units or M1/E2 mixing ratios for transitions between states of spin differing by one, combined with the static magnetic dipole moment of a state of the band. Of these, the static dipole moments are the rarest. In addition, knowledge of the intrinsic quadrupole moment Q$_{\rm 0}$ of the band states is required.  

\section{\label{exp}Experimental details}
\subsection{Nuclear orientation of the 37/2$^-$, 51.4 m, K-isomer in $^{\rm 177}$Hf}
At the ISOLDE isotope separator facility, CERN, new results have been obtained on the magnetic moment of the 37/2$^-$ high spin K-isomer and on gamma transitions in its decay to the ground state by the method of on-line nuclear orientation combined with NMR \cite{book}. The 1.4 GeV proton beam from CERN's PSB synchrotrons was incident upon a mixed tantalum/tungsten foil target and separation of the  Hf isotopes was achieved by introducing CF$_{\rm 4}$ gas into the plasma ion source and extracting a beam of HfF$_{\rm 3}^{+}$ ions \cite{koester2007}. The ions, accelerated through 60 keV, impinged upon a pure iron foil soldered to the cold finger of the NICOLE on-line nuclear orientation system \cite{eder1990} maintained at temperatures down to about 12 mK. The HfF$_{\rm 3}^{+}$ ions disintegrated at the foil surface and the Hf activity was implanted into the Fe lattice.  By interrupting the implantation and allowing the dilution refrigerator to cool further, data were taken to about 6 mK as the activity decayed. The iron foil was magnetized by an applied magnetic field of 0.5 T. 

\begin{figure*}
\includegraphics[ width=1.0\textwidth]{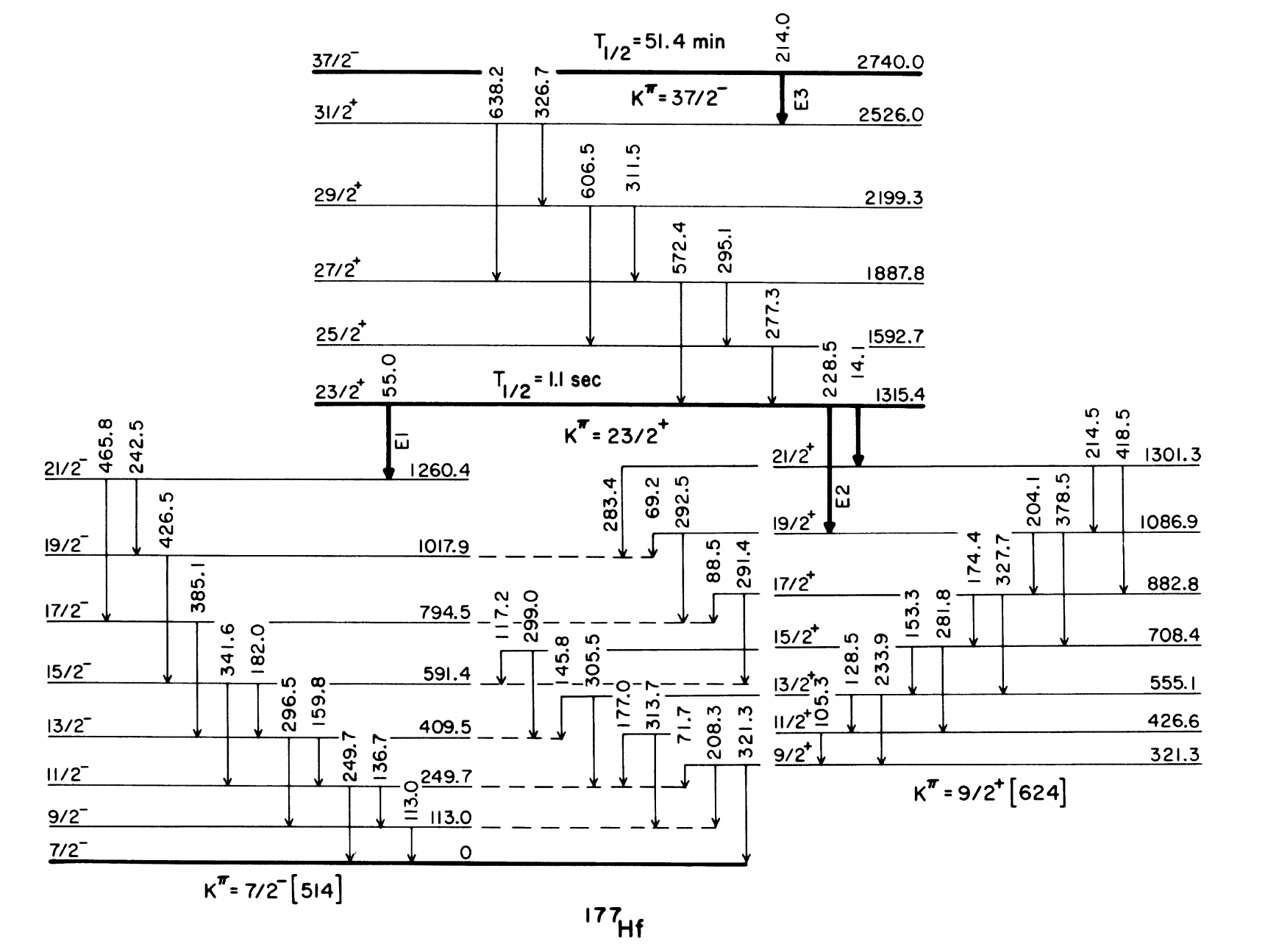}
\caption{\label{decay}$^{\rm 177}$Hf$^{m2}$ decay scheme. Taken from \cite{chu1972}.}
\end{figure*}

\begin{figure}
\includegraphics[width=9cm,height=7cm]{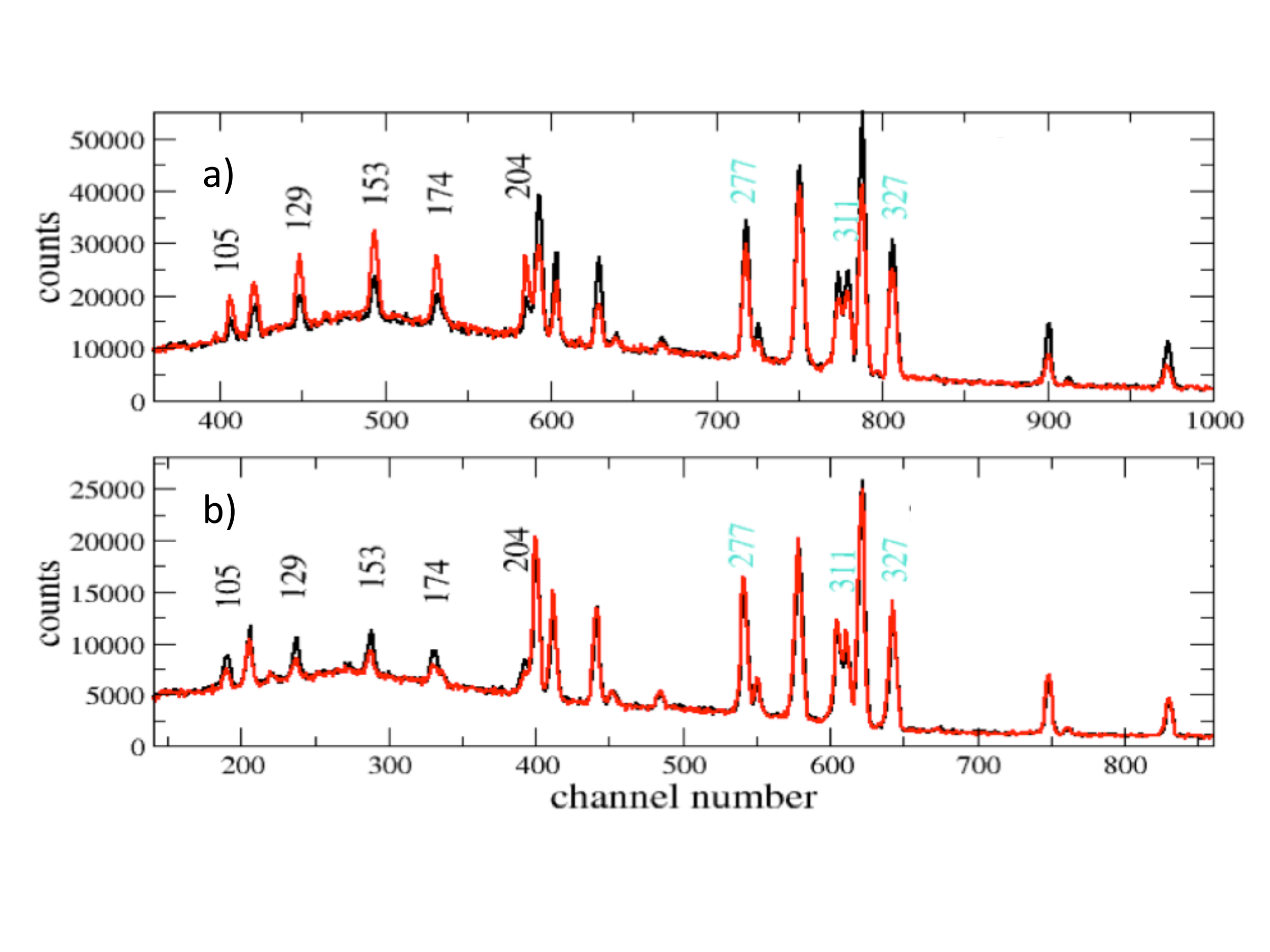}
\caption{\label{spectra} (color online) Warm and cold spectra from 0$^{\rm 0}$ and 90$^{\rm 0}$ detectors for two series of E2/M1 transitions. Black/dark energy labels - in band built on 9/2$^+$ [624] state - note increases at 0$^{\rm 0}$. Blue/pale energy labels - in band built on 23/2$^+$ three-quasi-particle isomer - note decreases at 0$^{\rm 0}$.}
\end{figure}

At mass A = 177 the Hf beam contains both the isomer $^{\rm 177}$Hf$^{\rm m2}$ (I$^{\rm \pi}$ = 37/2$^-$ T$_{\rm 1/2}$ = 51.4 m) and stable $^{\rm 177}$Hf ground state nuclei. The isomer activity made up about 1 \% of the beam. The decay scheme of $^{\rm 177}$Hf$^{\rm m2}$ is shown in Fig~\ref{decay}. Angular distribution parameters for all transitions were obtained from measurements, as a function of the foil temperature, of the gamma photo-peak count rates in two pairs of high resolution Ge detectors placed at angles of 0, 90, 90, and 180 degrees to the axis of polarization. The counts were normalized to rates measured with the activity un-oriented at a temperature close to 1 K. The foil temperature was obtained from measurements of the angular distributions of the 1333 keV and 1173 keV transitions in the decay of $^{\rm 60}$Co activity in a Co single crystal soldered to the back of the cold finger. Since, for the $^{\rm 60}$Co$\underline{Co}$ system, all orientation parameters are known, these measurements act as an accurate thermometer in the region between about 50 mK and 5 mK.
High degrees of polarization were achieved, produced by the large hyperfine field experienced by the Hf activity in the iron lattice. Examples of cold, oriented, gamma spectra compared with warm (1 K, un-oriented) spectra in 0$^{\rm 0}$ and 90$^{\rm 0}$ detectors are shown in Fig.~\ref{spectra}, in which the strongly differing anisotropies in transitions between levels in different bands of the decay are clearly seen. In addition to gamma transition angular distribution measurements, NMR of the implanted $^{\rm 177}$Hf$^{\rm m2}$ parent state activity was sought by exposing the implanted nuclei to an RF field, transverse to the axis of polarization, produced by a simple two-turn coil.
\vskip 2cm

\subsection{Hyperfine interaction and nuclear spin-lattice relaxation time estimates.}
Implanted nuclei have a high probability of coming to rest in regular lattice sites and the well-defined hyperfine interaction allows observation of Nuclear Magnetic Resonance by the disturbance caused to the gamma decay angular distribution by resonant absorption of RF energy. However even though the foil lattice may be at millikelvin temperature the nuclei are implanted un-oriented and it takes a time, related to the nuclear spin-lattice relaxation time, T$_{\rm 1}$ for nuclear polarization to be established \cite{shaw1989}. Both the resonant NMR frequency, $\nu_{\rm res}$ and T$_{\rm 1}$ depend upon the strength of the hyperfine structure splitting of the state undergoing resonance and/or orientation, h$\nu_{\rm res}$ = $\mu$B$_{\rm hf}$/I.

The nuclear moment of the 37/2$^-$ isomer can be estimated to be $\sim$7 $\mu_{\rm N}$ (see Sec.~\ref{moment} below) and the hyperfine field at Hf in Fe is B$_{\rm hf}$ = - 67.4(9) T (\cite{muto2004} improved using $^{\rm 175}$Hf moment 0.677(9) \cite{nieminen2002}). These values yield $\nu_{\rm res}$ in the region of 200 MHz. There have been no spin lattice relaxation time measurements for Hf isotopes in iron, however an empirical relationship, C$_{\rm K}$T$^{\rm 2}_{\rm int}$ = 1.4 x 10$^{\rm -4}$ sK$^{\rm 3}$, where T$_{\rm int}$ is the nuclear interaction strength, T$_{\rm int}$ = $\mu$B$_{\rm hf}$/kI, can be used to give estimates of the Korringa constant C$_{\rm K}$, usually found reliable to better than a factor of two \cite{shaw1989}. The resulting prediction is T$_{\rm 1}$ (37/2$^-$) $\sim$ 19 s at 10 mK which is much less than the isomer lifetime of 51.4 m, so full thermal equilibrium between the $^{\rm 177}$Hf$^{\rm m2}$ nuclei and the lattice will be established prior to decay. 

There remains the possibility that orientation may be perturbed by further interaction with the lattice during the lifetime of the 1.1 s 23/2$^+$ isomeric state at 1315 keV. Any such re-orientation would affect observed anisotropies in transitions below the isomer. However, an estimate of T$_{\rm 1}$ for the 23/2$^+$ isomer, based on an estimated 23/2$^+$  moment and the above T$_{\rm 1}$ (37/2$^-$) estimate gives the value T$_{\rm 1}$ (25/2$^+$) $\sim$ 8 s at 10 mK. This is substantially longer than the life time of the isomer, indicating that any re-orientation effect will be small.

\subsection{Nuclear orientation of the 8$^-$, 5.5 h, 1142 keV, state in  $^{\rm 180}$Hf}
At mass 180 the Hf beam has a strong isomeric $^{\rm 180}$Hf component which decays by mixed E1/M2/E3 transitions to 8$^+$ and 6$^+$ states in the ground state band. These states decay by a sequence of pure E2 transitions which show strong anisotropy as has been reported previously \cite{stone2007}. Earlier measurements of the magnetic dipole moment of this isomer have not been of high accuracy. K\"{o}rner et al. \cite{korner1971} reported 8.7(10) $\mu_{\rm N}$ by a M\"{o}ssbauer method and Krane et al \cite{krane1976} 9.0(9) $\mu_{\rm N}$ by an integral nuclear orientation experiment on sources of $^{\rm 180m}$Hf in HfZrFe$_{\rm 2}$ alloys. Since NMR on the implanted activity would provide a more precise result for the moment, resonance was sought with the activity implanted into an Fe foil. Unfortunately, at the high frequency at which resonance is expected for this isomer (in the region of 520 MHz) the RF line into the Nicole dilution refrigerator shows strong standing wave resonances which result in non-nuclear-resonance heating of the sample and render true resonance unobservable. In a second experiment the activity was implanted into a pure Ni foil in which the magnetic hyperfine field at Hf nuclei is much reduced. Resonance was observed at a frequency of 92.2(2) MHz in an applied field of 0.10 T. The result is discussed further below.

\section{\label{res}Analysis of Results}
\subsection{Gamma transition anisotropies}
Photo-peak counts for each transition in each spectrum for each gamma detector were obtained using the fitting code DAMM.  Counts taken with the sample at temperatures with appreciable orientation were normalized by taking ratios with counts from unoriented, warm (1 K), spectra. The gamma transition anisotropies, $A$, were obtained for each of the two pairs of detectors using the expressions
\begin{equation}
W(\theta, T) = N(\theta, T)/N(\theta,{\rm warm})
\end{equation}
and 
\begin{equation}
A = [W(0^{\rm 0},T)/W(90^{\rm 0},T)  -  1]
\end{equation}
The anisotropies were analysed further using the standard nuclear orientation formalism \cite{kranebook}
\begin{equation}
W(\theta,T)=1+f\sum_{\lambda}{B_{\lambda}U_{\lambda}A_{\lambda}Q_{\lambda} P_{\lambda}(cos \theta)}.
\end{equation}
Here $f$ is the fraction in good sites. The B$_{\lambda}$ factors describe orientation of the parent, isomeric, state and depend  upon the strength of the hyperfine interaction, the spin and the temperature. The U$_{\lambda}$ factors are calculated for each state below the parent and require knowledge of the spins of any intervening states, the intensities of all transitions between them and their multipolarity, with admixtures treated without regard to phase. The A$_{\lambda}$  parameters describe the observed emission and depend upon the initial and final state spins and the multipolarity, treated with sensitivity to phase of any multipole mixing ratio $\delta$.  Q$_{\lambda}$  are correction factors which account for the finite solid angle subtended by the detectors and P$_{\lambda}$  are the associated Legendre Polynomials. Since gamma emission conserves parity, the summation has only even terms (up to $\lambda$ = 2L$_{\rm max}$, where L$_{\rm max}$ is the highest multipolarity in the observed emission). 

In all on-line nuclear orientation experiments it has been found that the angular distributions can be well described as consisting of two components, the first from nuclei coming to rest in undisturbed (good) lattice sites and subject to the full hyperfine interaction and the second un-oriented at all temperatures. The un-oriented fraction describes nuclei which are stopped in the thin oxide layer which is always present on the implantation foil surface or are stopped in other very disturbed regions which are not magnetically ordered.

\subsection{Analysis of the pure E2, $\Delta$I = 2, transitions: determination of the fraction in good sites}
\begin{figure}[t]
\includegraphics[width=9cm]{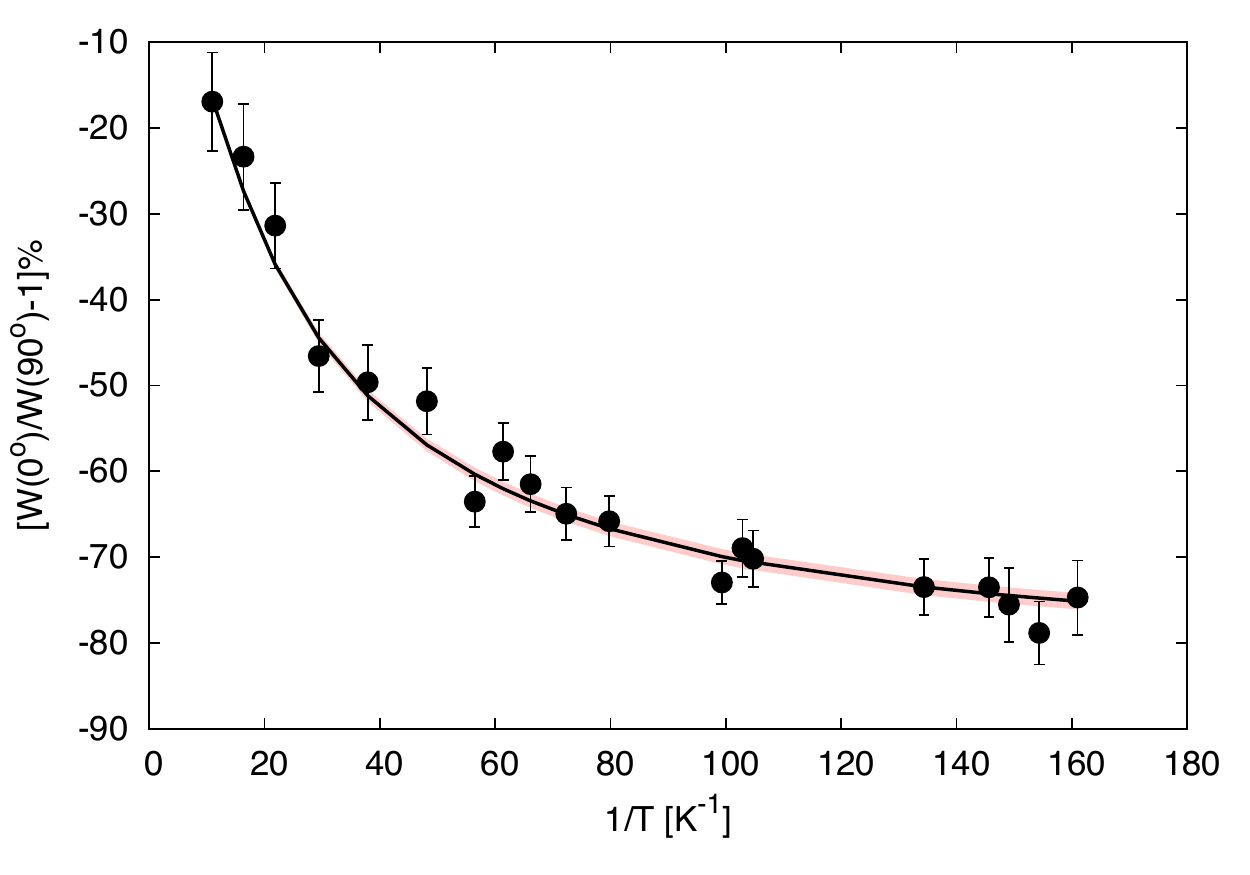}
\caption{\label{frac2}(color online) Anisotropy of the 606.5 keV 29/2$^+$ - 25/2$^+$ pure E2 transition in band built on 23/2$^+$ isomer. Fitted curve with shaded band is for $f$ = 0.786(12). }
\end{figure}

The first transition emitted by the 37/2$^{-}$ isomer is the 214 keV pure E3 transition. However this is an unresolved doublet with a mixed transition lower in the decay scheme so could not be used to evaluate the fraction $f$. The U$_\lambda$ values for this pure multipole transition are known, however, so $f$ can be obtained from the anisotropy of the pure E2, 638 keV, (31/2$^{+}$  - 27/2$^{+}$) transition. As the other parameters necessary to calculate U$_{\lambda}$ for lower states can all be extracted from the experimental data, values for $f$ can also be found from data on the other E2 transitions between pairs of lower states built on the 23/2$^{+}$ K-isomer. The results are given in Table~\ref{fraction}.  As an example, in Fig.~\ref{frac2} are shown the data for 606.5 keV transition between the 29/2$^{+}$ and 25/2$^{+}$ states  and calculation using the fit $f$ value, with its uncertainty.
\begin{table}
%\scriptsize
%\squeezetable 
\caption{\label{fraction}Fraction $f$ of implanted Hf nuclei in good sites as extracted from fitting E2 transitions with known decay parameters.}
\vspace{5pt} 
\begin{tabular}{llccc}
\hline
\multicolumn{2}{c}{Transition}& Energy  & Fitted fraction $f$ \\  \hline
I$_{\rm 1}$  &	   I$_{\rm 2}$	&	[keV]        &	     \\	\hline		
\multicolumn{4}{l}{Above the 23/2$^+$ isomer}  \\   \hline
31/2	         &  27/2  	&	638.2	&	0.755(7)    \\
29/2       	&  25/2	&	606.5	&	0.786(12)	\\
27/2 	&  23/2	&	572.4	&	0.768(12)	\\  \hline
\multicolumn{4}{l}{Below the 23/2$^+$ isomer}  \\   \hline
23/2 	&  19/2	&	228.5	&	0.773(6)	\\
21/2 	&  17/2	&	418.5	&	0.773(13)	&	\\
19/2 	&  15/2	&	378.5       &     0.773(6)            &       \\
15/2	        &  11/2	&	281.8	&	0.779(9)	&    \\   
13/2         &  9/2      &     233.9         &     0.802(21)   &    \\  \hline
\end{tabular}
\end{table}
\begin{figure}
\includegraphics[width=9cm]{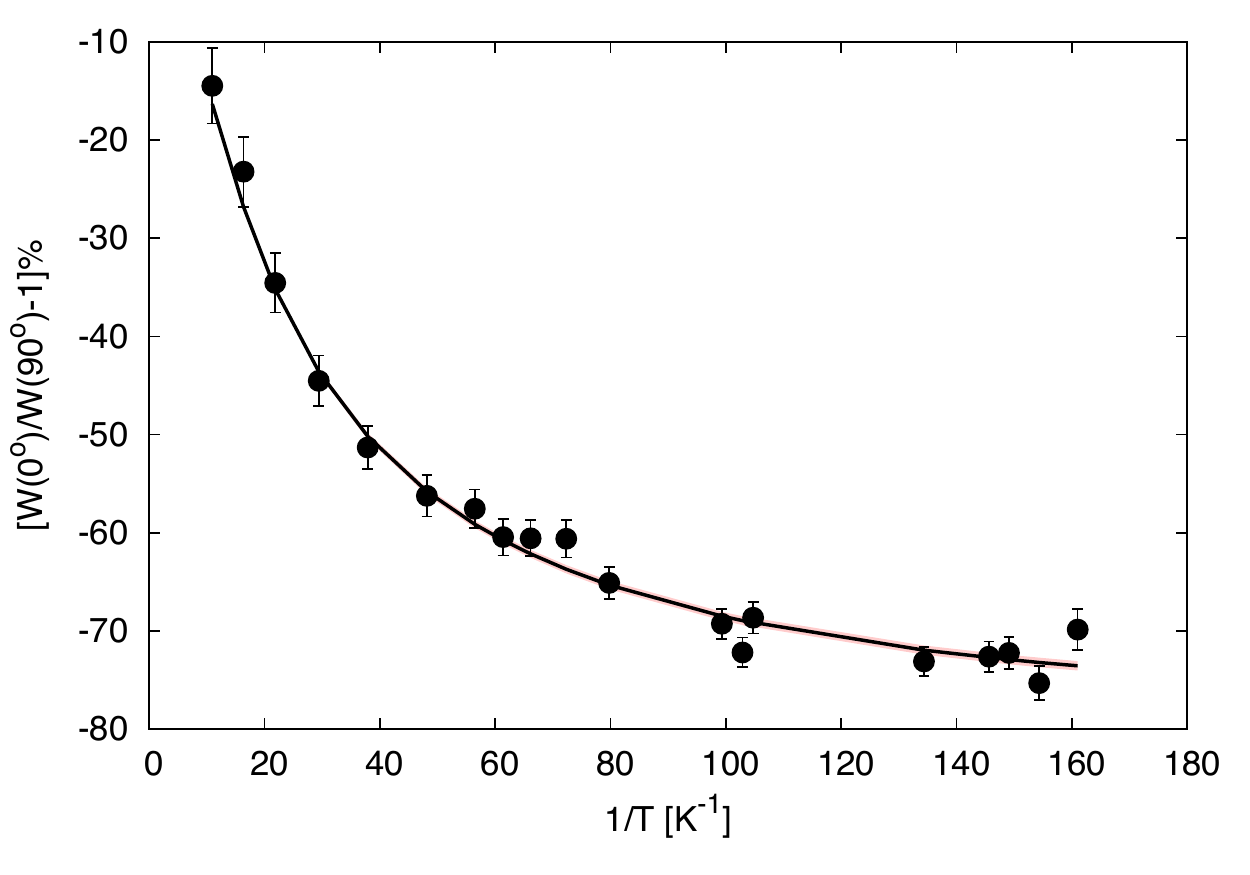}
\caption{\label{frac1}(color online) Anisotropy of the 228.5 keV 23/2$^+$ - 19/2$^+$ pure E2 transition from the 23/2$^+$ isomer to the band built on the 9/2$^+$ [624] band. Fitted curve with shaded band is for $f$ = 0.773(6). }
\end{figure}

\subsection{Re-orientation in the 23/2$^{+}$ isomeric state}
This state has lifetime 1.1 s and, although estimates given above suggest that the spin-lattice relaxation time T$_{\rm 1}$ of this state is considerably longer than the lifetime, this point requires further attention if anisotropies of gamma transitions between states below the isomer are to be quantitatively analysed. Data on anisotropies of the series of pure E2 transitions between states below the 23/2$^{+}$ isomer have all be treated to yield values of $f$ on the assumption that re-orientation may be neglected. The results are given in Table~\ref{fraction}. The close agreement between values of $f$ obtained from transitions above and below the isomer show clearly that the assumption is indeed valid. In Fig.~\ref{frac1} the data on the 228.5 keV transition between the 23/2$^{+}$ and 19/2$^{+}$ states are shown, with calculation using the fit $f$ value, with its uncertainty.
\begin{table}
\caption{\label{delta}E2/M1 mixing ratios in transitions from levels of $^{\rm 177}$Hf.}
\vspace{5pt} 
\begin{tabular}{lcccccc}
\hline
K      &  Level  &  E$_{\gamma}$  & $\delta$(InBeam)   &  $\delta$($^{\rm 177}$Lu) &  $\delta$($^{\rm 177}$Hf) &  (g$_{\rm K}$-g$_{\rm R}$)/Q$_{\rm 0}$ \\
       &   spin    &    [keV]      &   \cite{mullins1998} & \cite{krane1974a}  &     this work                   \\   \hline
37/2   &  41/2 &   420.9  &    0.74(23)        &                           &                 &    0.025(8)     \\
       &  43/2 &   440.0    &    0.74(16)        &                         &                 &    0.025(5)      \\  \hline
23/2   &  25/2 &   277.3    &                    &                         &      0.302(4)   &    0.069(1)  \\
       &  27/2 &   295.1  & 0.31(2)            &                           &                 &               \\               
       &  29/2 &   311.5  & 0.36(2)            &                           &      0.285(5)   &    0.071(2)  \\
       &  31/2 &   326.5  & 0.30(1)            &                           &      0.278(5)   &    0.071(2)  \\
       &  33/2 &   340.1  & 0.26(1)            &                           &                 &               \\
       &  35/2 &   351.9  & 0.21(1)            &                           &                 &               \\
       &  37/2 &   361.7  & 0.18(2)            &                           &                 &               \\
       &  39/2 &   369.4  & 0.20(2)            &                           &                 &               \\   \hline
9/2    & 11/2  &  105.3   &                    &   -0.36(4)                &        -0.23(4) &   -0.083(15)  \\
       & 13/2  &  128.4   & 0.38(1)            &   -0.37(6)                &       -0.34(3)  &   -0.055(5)   \\ 
       & 15/2  &  153.1   & 0.38(1)            &   -0.33(5)                &       -0.317(13)&   -0.061(3)   \\
       & 17/2  &  174.3   & 0.36(1)            &   -0.32(4)                &       -0.296(13)&   -0.065(3)   \\
       & 19/2  &  204.1   & 0.40(1)            &   -0.33(5)                &       -0.289(13)&   -0.070(4)   \\
       & 21/2  &  214.3   & 0.35(1)            &   -0.29(2)                &                 &                \\
       & 23/2  &  260.5    & 0.50(2)           &                           &                 &                \\
       & 25/2  &  241.8   & 0.40(2)            &                           &                 &                \\
       & 27/2  &  325.4   & 0.61(7)            &                           &                 &                \\
       & 29/2  &  249.4   & 0.27(2)            &                           &                 &                \\   \hline
\end{tabular} 
\end{table}

The data from both pairs of detectors were in very close agreement. An average value $f$ = 0.775(5) was adopted and used throughout the analysis.
\begin{figure}
\includegraphics[width=9.0cm]{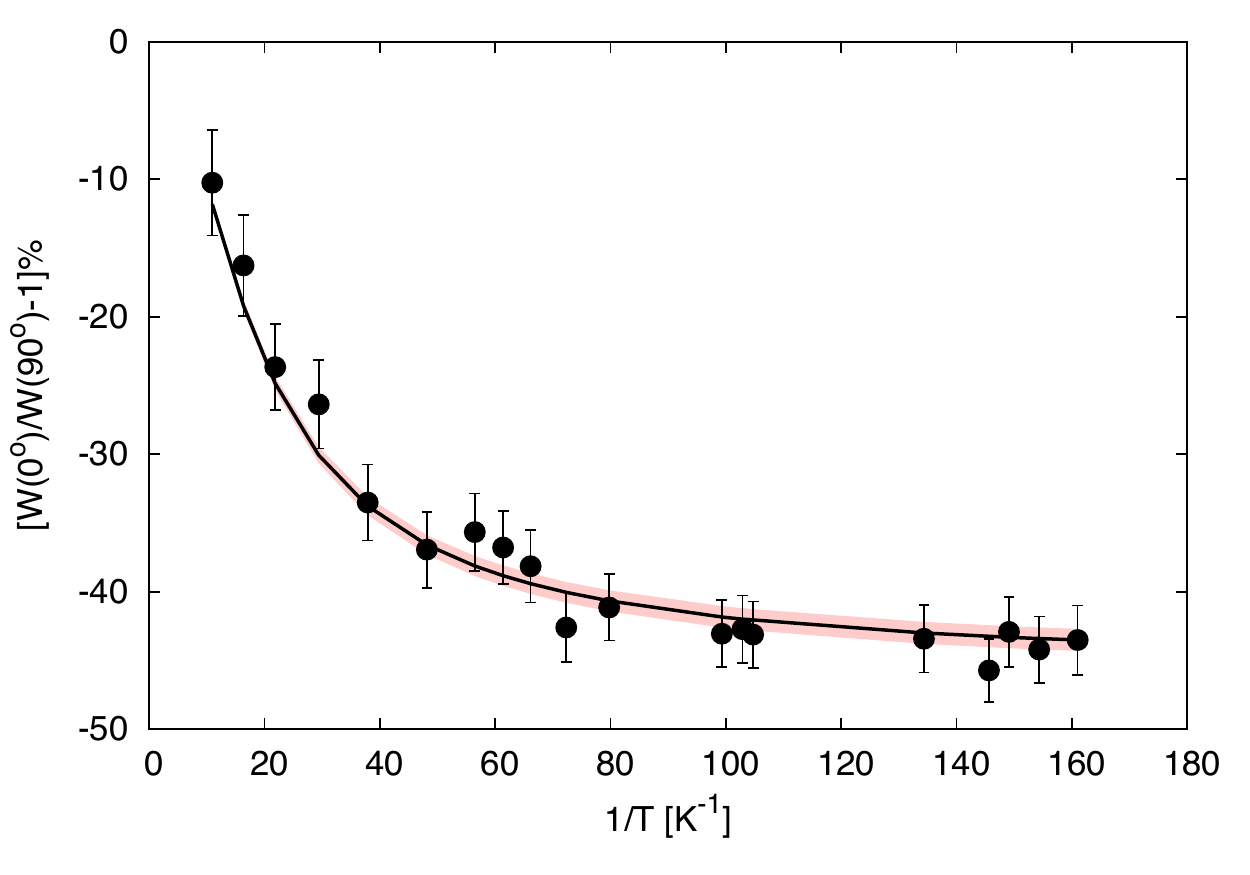}
\caption{\label{anis1}(color online) Anisotropy of the 277.3 keV 25/2$^+$ - 23/2$^+$ mixed E2/M1 transition in band built on 23/2$^+$ isomer. Fitted curve with shaded band is for $\delta$ [E2/M1] = +0.302(4). }
\end{figure}
\begin{figure}
\includegraphics[width=9cm]{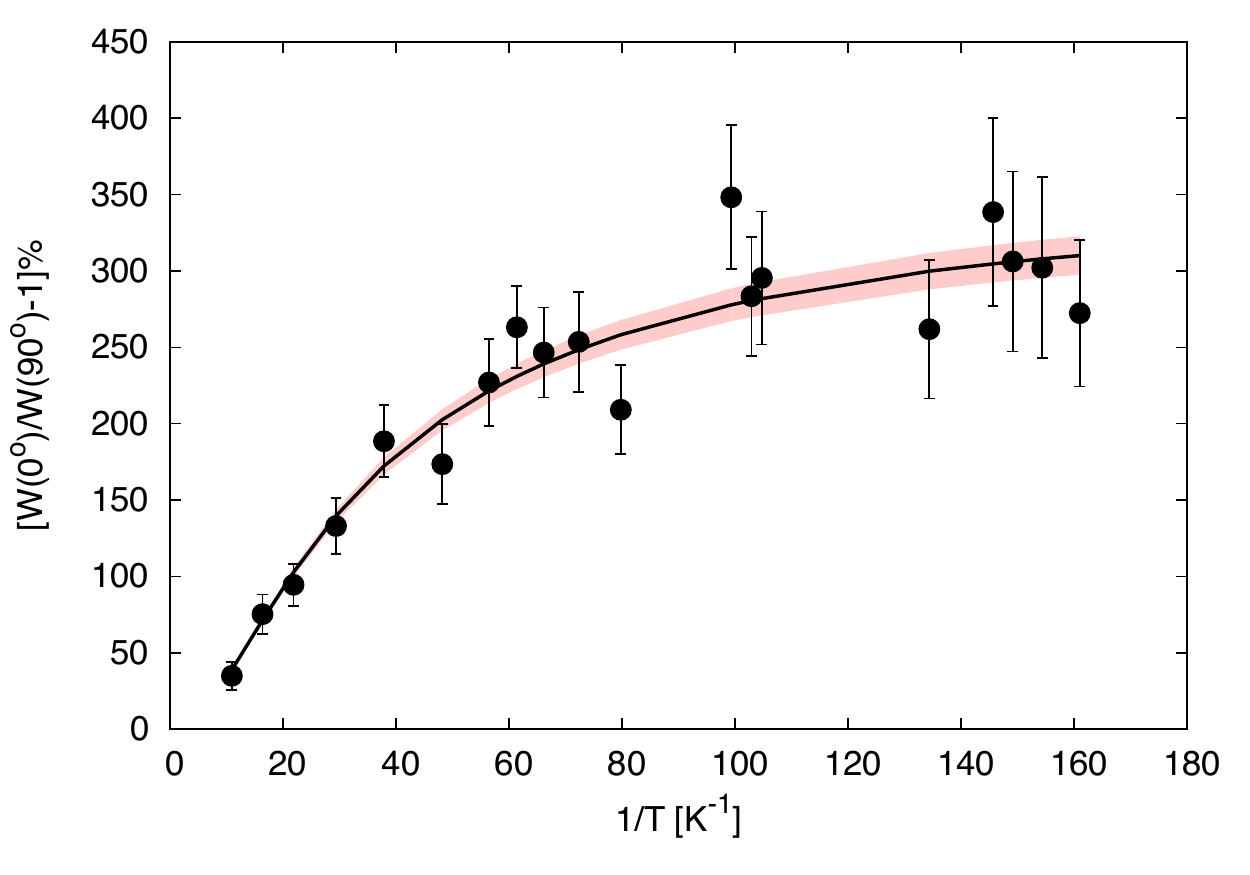}
\caption{\label{anis2}(color online) Anisotropy of the 153.1 keV 15/2$^+$ - 13/2$^+$ mixed E2/M1 transition in band built on 9/2$^+$ [624] band. Fitted curve with shaded band is for $\delta$ [E2/M1] = -0.317(13). }
\end{figure}
\subsection{Analysis of the mixed multipole $\Delta$I = 1 transitions: determination of the E2/M1 mixing ratios.}
Each state in the bands built upon the 23/2$^{+}$ multi-quasi-particle isomer and the 9/2$^{+}$ and 7/2 $^{-}$  single quasi-particle states decays both by a direct E2, $\Delta$I = 2, transition and a mixed E2/M1, $\Delta$I = 1, transition. The E2/M1 multipole mixing ratio, $\delta$, is a most useful quantity in establishing the parameters of the band, as described previously. For each of these transitions the anisotropies were analysed in a somewhat novel way to give the best fit values of $\delta$. Each data point in the temperature dependence (19 in all) was regarded as an individual experiment and fitted with $\delta$ as the only unknown. The resulting set of $\delta$Õs was treated statistically to give the best fit value and the error on this value. Results are shown in Table~\ref{delta}, compared to results from other investigations \cite{mullins1998, krane1974a}. The overall agreement is excellent and the new results have significantly smaller experimental uncertainties. Examples of the fits are shown in Figs. ~\ref{anis1} and \ref{anis2}. Fig.~\ref{anis1} shows data for the 277.3 keV transition between the 25/2$^{+}$ and 23/2$^{+}$ states in the 23/2$^{+}$ band. All mixed transitions in this band show similar negative anisotropies. Fig.~\ref{anis2} shows data for the 153.1 keV transition between the 15/2$^{+}$ and 13/2$^{+}$ states in the 9/2$^{+}$ band. All mixed transitions in this band show similar large positive anisotropies. The larger scatter and uncertainties for these transitions arise because they are measured on a large Compton-backscattered gamma intensity so that the small 90$^{\rm 0}$ counts at lower temperatures have large uncertainties (see Fig.~\ref{spectra}). 
\begin{figure}
\includegraphics[width=9.0cm]{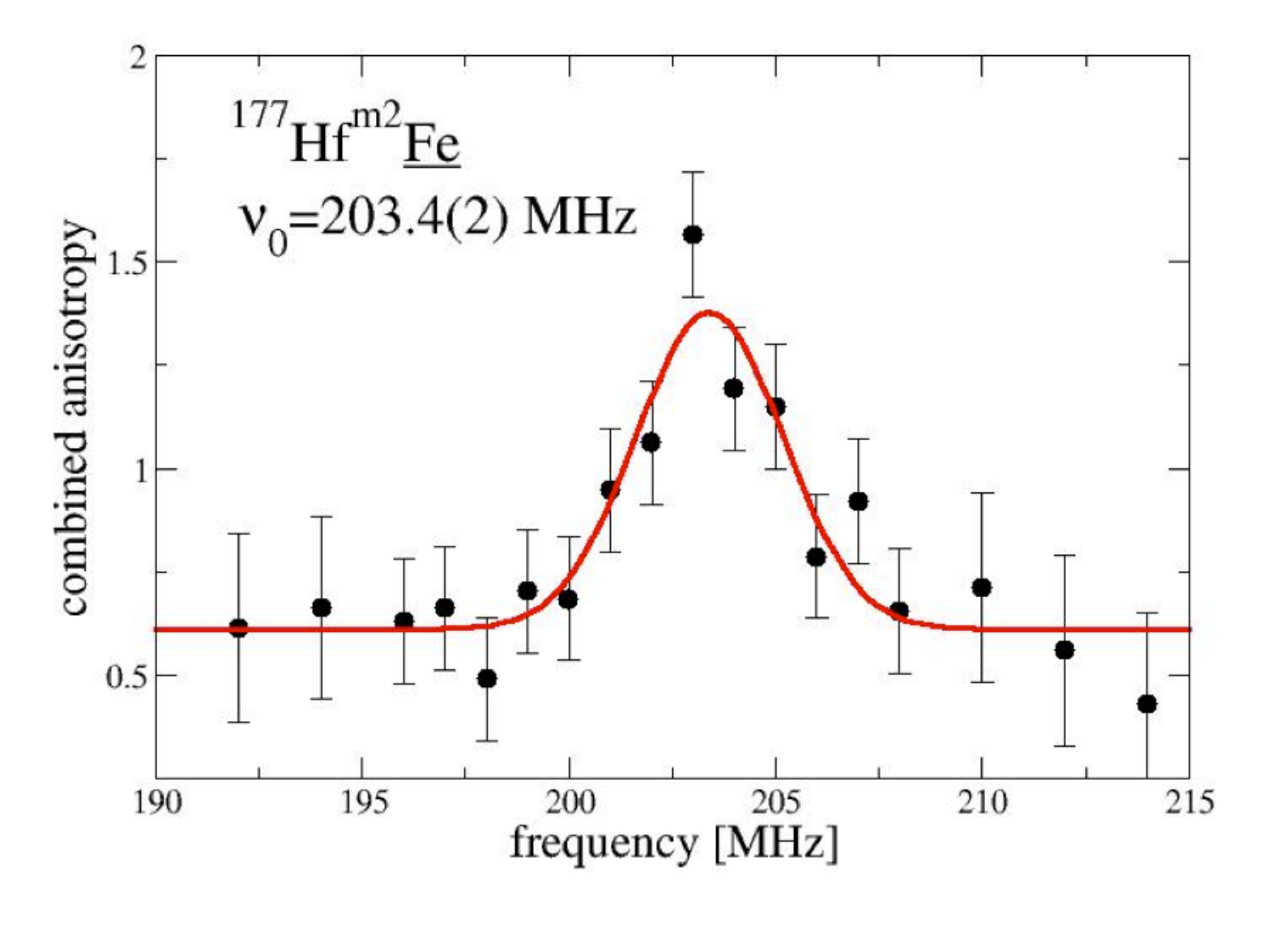}
\caption{\label{fig:res177}(color online) Resonance observed in a combined signal from the stretched E2 transitions in the decay of the 37/2$^-$ isomer in $^{\rm 177}$Hf.}
\end{figure}
\subsection{\label{moment}The magnetic moment of the 37/2$^{-}$ isomer in $^{\rm 177}$Hf}
The search for NMR of the 37/2$^{-}$ isomer was carried out with RF swept between 180 and 230 MHz in 1 MHz steps with RF modulation of 1 MHz. A positive signal was found, as shown in Fig.~\ref{fig:res177}, with centre frequency
\begin{equation}
\nu_{\rm res} (^{\rm 177}{\rm Hf}^{\rm m2} \underline{Fe})  =  {\rm 203.4(2)\mbox{}MHz}.
\end{equation}
This frequency, combined with the known hyperfine field -67.4(9) T, adjusted for the external field of 0.100 T applied to the Fe foil during the NMR measurements, yields the magnitude of the magnetic moment of the 37/2$^{-}$ isomer as
\begin{equation}
|\mu|(^{\rm 177}{\rm Hf}^{\rm m2},\mbox{} 37/2^{-}, \mbox{} {\rm 51.4 m})  = {\rm  7.33(9)} \mbox{ }{\rm \mu_{\rm N}}
\end{equation}
Note that any hyperfine anomaly between the isotope $^{\rm 175}$Hf, on which the hyperfine field measurement was made, and $^{\rm 177}$Hf$^{\rm m2}$ is not expected to exceed 0.1\% and has been neglected.
\begin{figure}
\includegraphics[width=9cm]{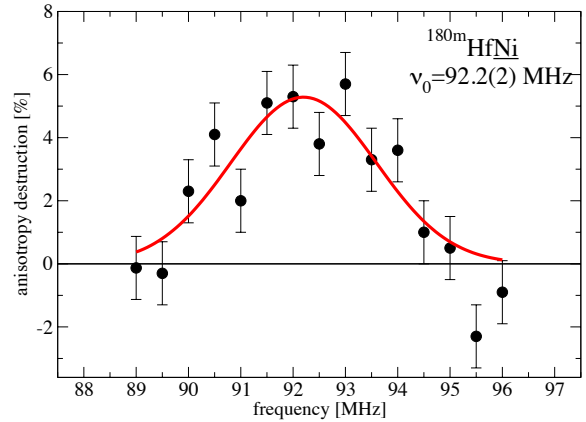}
\caption{\label{fig:res180}(color online) Resonance observed in percentage destruction of anisotropy of the 8$^-$ isomer in 
$^{\rm 180}$Hf.}
\end{figure}

\subsection{The magnetic moment of the 8- isomer in $^{\rm 180}$Hf}
Extraction of the magnetic moment from the observed resonance (Fig.~\ref{fig:res180}) requires knowledge of the hyperfine field acting on Hf nuclei in the Ni lattice. Several previous values for this field have been reported, none of them of the necessary accuracy. The best would appear to be the M\"{o}ssbauer study by Aggarwal et al. \cite{aggrawal1983}  who gave the value B$_{\rm hf}$(Hf$\underline{Ni}$) = 11.8(2.6) T. Using this value the observed $^{\rm 180m}$Hf resonance yields the moment of the isomer to be 8.3(1.8) $\mu_{\rm N}$. Other, lower, reported Hf$\underline{Ni}$ fields give unrealistically high values for the nuclear moment. Whilst the hyperfine fields at some neighbouring elements have better measurements in Nickel they do not allow an improved estimate for the field at the Hf nucleus. Clearly as it stands the present value is no improvement on the earlier results and the observed NMR result awaits an improved field value. This can be obtained, for example, by the observation of NMR in $^{\rm 177m2}$Hf activity in Ni where the frequency should be in the region of 35 MHz.
\begin{table*}
%\scriptsize
%\squeezetable 
\caption{\label{states}Adopted g$_{\rm K}$ value for individual quasi-particle states. Experimental data taken from \cite{stone}.} 
\vspace{5pt} 
\begin{tabular}{ccll}
\hline
Quasi-particle state	&	Adopted g$_{\rm K}$  	& Basis of adopted value\\  \hline
Protons      &                 &      \\    \hline
7/2+[404]	&	0.765(25)	&	moments of 7/2+ ground states in $^{\rm 175,177,179,181}$Ta  \\
9/2-[514]	&	1.37(3)	&		moment of 9/2- 6 keV state in $^{\rm 181}$Ta  \\
Neutrons     &                &   \\   \hline
5/2-[512]	&	-0.48(2)	&	moment of 5/2- ground state of $^{\rm 175}$Hf  \\
7/2-[514]	&	0.206(14)	&	moment of 7/2- ground state of $^{\rm 177}$Hf  \\
9/2+[624]	&	-0.239(11)	&	moment of 9/2+ ground state of $^{\rm 179}$Hf   \\  \hline
\end{tabular}
\end{table*}
\begin{figure}
\includegraphics[width=9.0cm]{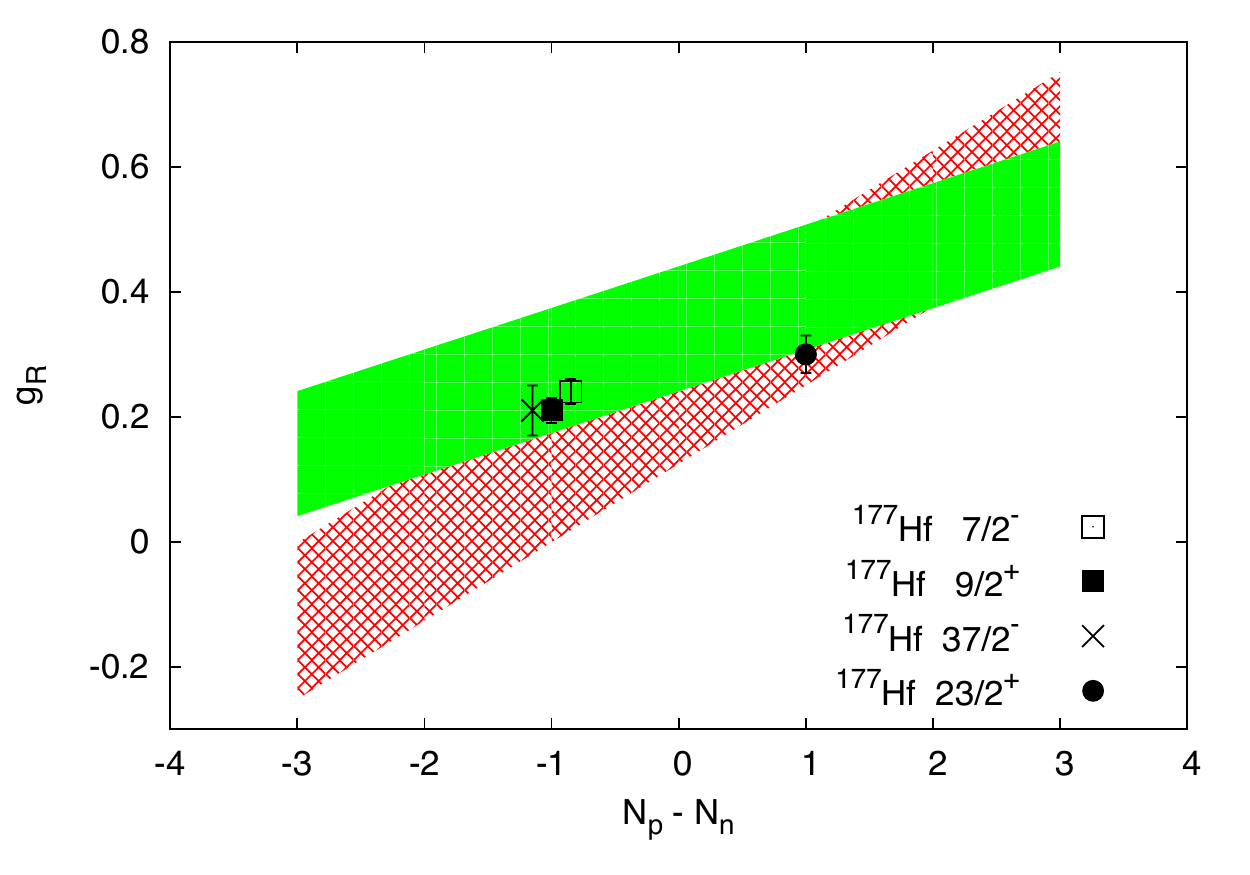}
\caption{\label{grplot}(color online) Plot of the collective g$_{rm R}$ as a function of $\Delta$ = N$_{\rm p}$ - N$_{\rm n}$. For ease of viewing the three data points at -1 have been slightly off-set along the x-axis. For discussion see text and Ref.~\cite{stone2013}.}
\end{figure}
\begin{table*}
%\scriptsize
%\squeezetable 
\caption{\label{tab:add}g$_{\rm R}$ values extracted from band-head moments and band spectroscopic data. Q$_{\rm o}$ taken as 7.2 eb throughout analysis.
}
\vspace{5pt} 
\begin{tabular}{lrrcccccc} \hline
Isotope  &	State  &  Energy  &  N$_{\rm p}$  & N$_{\rm n}$  & $\Delta$    &   Moment  & Spectroscopy  &	g$_{\rm R}$   \\	
         &         &  [keV]   &            &           & = N$_{\rm p}$-N$_{\rm n}$   &   Reference  &   Reference   &      \\   \hline	
$^{\rm 177}$Hf  &	37/2-	 & 2740   &  2  & 3     &	-1    &   this work             &\cite{mullins1998} &  0.21(4) \\
$^{\rm 177}$Hf  &	  7/2- &   0 &  0  & 1     &	-1	&  \cite{buttgenbach1973} & \cite{mullins1998}   &  0.24(2)  \\
$^{\rm 177}$Hf  &	 9/2+	 &  321&  0  & 1     &	-1    &   \cite{hubel1969}         &    this work	          &  0.21(2)  \\
$^{\rm 177}$Hf  &	23/2+	 & 1316   &  2  & 1     &	+1    &   additivity             &	this work           &  0.30(3) \\
\end{tabular}
\end{table*}
\section{\label{dis}Discussion}

The variation in g$_{\rm R}$ found in the analysis of this work (see below) prompted, in part, a comprehensive survey of nearly one hundred high-K bands in isotopes in this region \cite{stone2013}. The results have been recently published and draw two significant conclusions: that the assumption of additivity of quasi-particle contributions is a sound basis for estimation of the quasi-particle g-factor g$_{\rm K}$ in multi-quasi-particle isomers in the majority of cases and that, based on analysis of the mixing ratio and branching ratio data, there is a wide and systematic variation of the collective g-factor g$_{\rm R}$ which depends upon the 
quasi-particle make-up of the band head isomer. Underlying ideas concerning pairing and blocking in these isomers are discussed in the paper.
In this section we review the extent to which the experimental results found in this work are consistent with the two conclusions described.

\subsection{Additivity of single quasi-particle g$_{\rm K}$ factors to form g$_{\rm K}$ for multi-quasi-particle states}

In Table~\ref{states} we give the g$_{\rm K}$ factors adopted in the broad survey for those single-quasiparticle states which are relevant to the two isomers $^{\rm 177m2}$Hf and $^{\rm 180m}$Hf for which the magnetic moments have been measured in this work. The g$_{\rm R}$ factor makes a small contribution to the total magnetic moment since it is multiplied by the factor I/(I+1) as compared to I$^{\rm 2}$/(I+1) for g$_{\rm K}$ (see Eq.~\ref{gk}) and the spins of the two isomers are high, 37/2 and 8, respectively. Taking g$_{\rm R}$ as 0.29(5) and assuming additivity, as described by Eq.~\ref{gks} yields predicted moments of these isomers of 7.28(13) $\mu_{\rm N}$ and 8.12(15) $\mu_{\rm N}$ to be compared to the measured values, 7.33(9) $\mu_{\rm N}$ and 8.3(18) $\mu_{\rm N}$ Both results show agreement with the predictions based on additivity, consistent with the findings in \cite{stone2013}. 

\subsection{The variation of g$_{\rm R}$}
Taking average values of the parameter (g$_{\rm K}$ - g$_{\rm R}$/Q$_{\rm 0}$) with Q$_{\rm 0}$ = 7.2 $e$b and g$_{\rm K}$ assuming additivity in the multi-quasiparticle states, values of g$_{\rm R}$ were extracted from this work for bands built on the 9/2+, 23/2+ and 37/2- states in $^{\rm 177}$Hf.  Results are given in Table~\ref{tab:add} which also includes g$_{\rm R}$ for the band built on the 7/2- ground state of $^{\rm 177}$Hf.
We reproduce in Fig.~\ref{grplot} the evidence of variation of g$_{\rm R}$ as a function of the difference, N$_{\rm p}$ - N$_{\rm n}$, of the numbers of unpaired quasi-protons and quasi-neutrons in the isomer reported in Ref.~\cite{stone2013}, showing the broad band ranges which, it is suggested, are caused by variation in the contributions from breaking specific quasi-particle pairs and the individual results from Table~\ref{tab:add}.

\section{Conclusions}
This paper reports new measurements of the magnetic properties of the nucleus $^{\rm 177}$Hf and $^{\rm 180}$Hf and their high K isomers.The magnetic moments of the 37/2- isomer in $^{\rm 177}$Hf and the 8$^{\rm -}$ isomer in $^{\rm 180}$Hf provide valuable additional evidence for the validity of additivity in estimating the quasi-particle g-factor g$_{\rm K}$ in these deformed nuclei. Values of the collective g-factor, g$_{\rm R}$, obtained from detailed gamma transition anisotropy measurements, are shown to be fully consistent with the recently revealed systematic dependence of this parameter upon the quasi-particle make-up of the bands involved.

\section{Dedication}
We dedicate this paper to the memory of our colleague Suguru Muto, a most valuable member of the team who performed this work, who died in early 2013 after a long battle with cancer.
 
\section{Acknowledgements}
We thank Ajay Deo for helpful assistance with the experiment. The research was supported by the US DOE Office of Science, the Ministry of Education and Science of Serbia (project No. 171002), the European Commission via the ENSAR project, SNRS and IN2P3 (France), JSPS KAKENHI (Grant Number 23540337) (Japan) and STFC (UK).

\end{document}